\documentclass{aa} 
\usepackage{graphicx}
\usepackage{txfonts}
\usepackage{xcolor} 
\usepackage{array} 
\begin{document} 

\title{BAO angular scale at $z_{\mbox{\footnotesize eff}} = 0.11$ with the SDSS blue galaxies
}



\author{E. de Carvalho\inst{1,2}\thanks{edfilho@uea.edu.br}
\and
A.  Bernui\inst{2}
\and
F. Avila\inst{2}
\and
C. P. Novaes\inst{3}
\and
J. P. Nogueira-Cavalcante\inst{2}
           }

\institute{
Centro de Estudos Superiores de Tabatinga, Universidade do Estado do Amazonas, 
69640-000, Tabatinga, AM, Brazil \\
\and
Observat\'orio Nacional, Rua General Jos\'e Cristino 77, S\~ao Crist\'ov\~ao, 
20921-400 Rio de Janeiro, RJ, Brazil \\
\and
Instituto Nacional de Pesquisas Espaciais, Av. dos Astronautas 1758, 
Jardim da Granja, S\~ao Jos\'e dos Campos, SP, Brazil
              }


\abstract
{} 
{We measure the transverse baryon acoustic oscillations (BAO) signal in the local Universe 
using a sample of blue galaxies from the Sloan Digital Sky Survey (SDSS) survey as a cosmological tracer.} 
{The method is weakly dependent on a cosmological model and is suitable for 2D
analyses in thin redshift bins to investigate the SDSS data in the interval $z \in [0.105,0.115]$.} 
{We detect the transverse BAO signal 
$\theta_{\mbox{\sc bao}} = 19.8^{\circ} \pm 1.05^{\circ}$ at 
$z_{\mbox{\tiny eff}} = 0.11$, with a statistical significance of $2.2\,\sigma$.
Additionally, we perform tests that confirm the robustness of this angular BAO signature. 
Supported by a large set of log-normal simulations, our error analyses include statistical and 
systematic contributions.
In addition, considering the sound horizon scale calculated by the Planck Collaboration, 
$r_s^{Planck}$, and the $\theta_{\mbox{\sc bao}}$ value obtained here, we 
obtain a measurement of the angular diameter distance 
$D_A(0.11) = 258.31 \pm 13.71\, h^{-1}$Mpc. 
Moreover, combining this $\theta_{\mbox{\sc bao}}$ measurement at low redshift with other 
BAO angular scale data reported in the literature, we perform statistical analyses for the 
cosmological parameters of some Lambda cold dark matter ($\Lambda$CDM) type models. 
}
{}

\keywords{Cosmological parameters -- Large-scale structure of Universe -- Observations}
               
\maketitle

\section{Introduction}\label{sec1}

Embedded in the 3D distribution of cosmic luminous matter are 
geometrical signatures from the primordial baryon acoustic oscillations 
\citep[BAO;][]{PeeblesYu,SunyaevZeldovich,BondEfstathiou87,Eisenstein05,Cole05}. 
They can be statistically revealed in large-scale and numerically dense astronomical surveys 
and are used as a standard ruler to measure our distance to the data region. 
These analyses are performed by studying different cosmological tracers from a variety of 
astronomical surveys, such as the Sloan Digital Sky Survey (SDSS), the 6dF Galaxy Survey, and 
the WiggleZ Dark Energy Survey~\citep{Alam17,Alam20,Beutler11,Blake11}. 
A set of precise distance measurements for several redshift values will unambiguously 
describe the dynamics of the universe~\citep{EisensteinHu98,Bassett,Eisenstein07}.

The BAO distance measurements are obtained using two-point statistics in at least two ways. 
The first approach, based on the 3D information, assumes a fiducial cosmology to transform 
the redshift of each cosmic object into its radial distance, and with the two angular coordinates 
measured in the survey, the comoving distance between all possible pairs is calculated to 
construct the two-point correlation function (2PCF). 
The BAO signal obtained with this approach determines the sound horizon scale at the end 
of the baryon drag epoch, $r_s$, and the spherically averaged distance 
$D_V$~\citep{Beutler11,Blake11,Alam17,Abbott18}. 
The second approach uses 2-dimensional (2D) information: 
the data in a redshift shell are projected on the celestial sphere. With the two angular 
coordinates of each cosmic object, the angular separation between pairs is then calculated and the two-point angular correlation function (2PACF) is calculated,  where the BAO angular scale 
provides a measure of the angular diameter distance $D_{\!A}$ if $r_s$ is known. 
To minimize projection effects that would affect this measurement, the data should be in a thin 
redshift shell~\citep{Sanchez11,Carnero12,Carvalho16}.

In addition to the advantages and disadvantages{\it } of each approach, the 2D method is a quasi 
model-independent procedure, with a weak dependence on the fiducial cosmology that we explained below. We adopt it here to measure the BAO angular scale 
$\theta_{\mbox{\sc bao}}$. 
The 2D approach was not widely applied to early data releases because the number density 
of cosmic objects was not high enough to provide a good BAO signal-to-noise ratio (S/N) in thin redshift 
shells. 
However, the current data releases have suitably increased this quantity. 
Several studies reported 2D BAO measurements using luminous red galaxies (LRG) and 
quasar samples at several redshifts~\citep{Sanchez11,Carnero12,%
Salazar17,Carvalho16,Abbott18,Edilson18}. 
The present work extends these analyses with a 2D BAO measurement at low redshift, 
$z_{\mbox{\tiny eff}} = 0.11$ from an unusual cosmological tracer, the SDSS 
sample of blue galaxies~\citep{SDSS,Avila19}.

This work is organized as follows. 
Section~\ref{sec2} provides the details of the blue galaxy sample selection from the SDSS 
data set, the generation of the random catalog, and the simulations we used.
Section~\ref{sec3} describes the statistical tools employed in the 2D clustering analyses.
In Sect.~\ref{sec4} we describe our main results, giving details on the 2D analyses and our 
estimate of the BAO angular scale, while in 
Sect.~\ref{sec5} we present the discussion of our results and final conclusions.

\section{Data description}\label{sec2}

\subsection{Blue galaxy sample and random catalogs}

\begin{figure}
\hspace*{-0.0cm}\includegraphics[width=1.0\columnwidth]{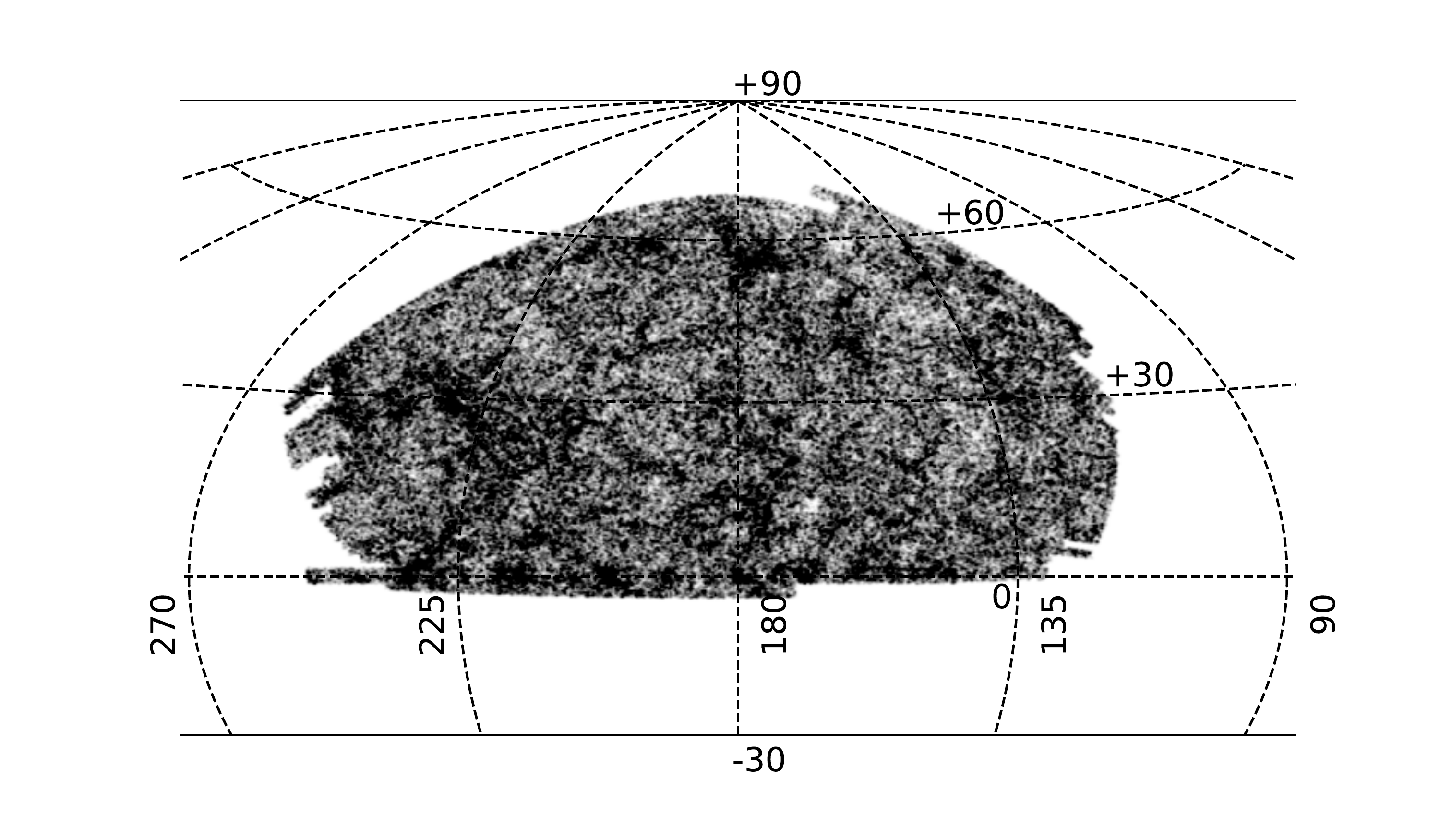}
\caption{Sample of SDSS blue galaxies in equatorial coordinates J2000 (in degrees).}
\label{fig:SDSSsample}
\end{figure}

We used the sample of blue star-forming galaxies analyzed in \cite{Avila19}.
The selected data are part of the twelfth public data release, DR12, of the SDSS collaboration 
\citep{Alam15}. 
We considered the low-redshift SDSS blue galaxies displayed in the north galactic cap with the 
footprint observed in Fig. \ref{fig:SDSSsample}, covering an area of $\sim$~7,000 deg$^2$.

To optimize between sample variance and shot noise, the galaxy field has to be weighted. 
To do this, we assigned weights to each galaxy based on the average local density in the analyzed 
region by using the Feldman-Kaiser-Peacock (FKP) weights~\citep{FKP}. This is a 
scale-independent weighting that depends on redshift, 
$\mbox{w}_{\mbox{\footnotesize\sc fkp}}(z) \,=\, 1 / (1 + n(z)P_{0})$, 
where $P_{0}$ is the amplitude of the power spectrum and $n(z)$ is the number density 
of galaxies. 
We used $P_{0} \simeq 10,000 \,h^{-3}$Mpc$^3$, the power amplitude relevant to the BAO 
signal $k \approx 0.15 h$Mpc$^{-1}$~\citep{Eisenstein05,Beutler11,Carter18}. 
Therefore the effective redshift of our sample, $z_{\mbox{\tiny eff}}$, 
calculated with the FKP galaxy weights, 
$\mbox{w}_i \equiv \mbox{w}(z_i) = \mbox{w}_{\mbox{\footnotesize\sc fkp}}(z_i)$, is
obtained through \citep[see, e.g.,][]{Carter18}
\begin{eqnarray}
z_{\mbox{\tiny eff}} = 
\frac{\sum_{i=1}^{N_g} \mbox{w}_i \, z_i}{\sum_{i=1}^{N_g} \mbox{w}_i} \, ,
\end{eqnarray}
where $N_g$ is the total number of galaxies in the sample.

We searched for a statistically significant angular BAO detection at the lowest redshift. 
After analyzing bins with a large number of galaxies (to minimize the statistical noise) that are located 
in a thin redshift bin (to minimize the nonlinear contributions due to the projection effect, see, 
e.g.,~\cite{Sanchez11}), we selected the data sample contained in the thin redshift bin 
$0.105 \le z \le 0.115$, with $\delta z = 0.01$ and $z_{\mbox{\tiny eff}} = 0.11$. 
This bin contains $N_g = 15,942$ blue galaxies. 

We would like to point out that the comoving volume survey containing these $N_g = 15,942$ 
blue galaxies is $V \simeq 0.0063 \,(\text{Gpc}/h)^3$. 
This seems to be a small volume in which to look for a transverse BAO signal, but 
the important parameter for these analyses is the number density $n$. 
For this sample of blue galaxies, the number density is 
$n = N_g / V \simeq 2.5 \times 10^6 (h/\text{Gpc})^3$. 
For comparison,  
$n_{\text{\,LRG}} = 10^4 - 10^5  \,(h/\text{Gpc})^3$ for the SDSS LRG sample analyzed in the 3D BAO detection~\citep{Eisenstein05}, or 
$n_{\text{quasars}} \simeq 7.3 \times 10^3  \,(h/\text{Gpc})^3$ for the 2D BAO detection 
using a sample of SDSS quasars~\citep{Edilson18}.

The random catalogs are an important ingredient in our analyses. 
They are necessary to extract the BAO features from the data. For this, they must have 
 properties in common to those observed in the SDSS blue galaxy catalog. 
We produced 50 random catalogs for the 2D analyses (with $N_{sim} \simeq 16,000$ 
in each catalog) with Poisson-distributed objects~\citep{PeeblesHauser} sharing the 
observational features of the data set in analyses (i.e., the same number density and footprint sky area as the SDSS data). 
Our set of random catalogs was produced following the 
method described in the Appendix B of \cite{Edilson18}, where they were 
satisfactorily tested through a null test 
analysis \citep[see, e.g., Sect. 5 of][]{Landy-Szalay} to confirm 
that they do not introduce spurious signals.

\subsection{Log-normal simulations}\label{mocks}

To estimate the error bars of the 2PACF and the statistical significance of our results, 
we used the covariance matrix built from full-sky log-normal simulations that we produced with 
the \texttt{FLASK} code\footnote{\url{http://www.astro.iag.usp.br/~flask}}~\citep{Xavier16,%
Edilson20}. 
We generated a set of $1,000$ simulations for which we assume the Lambda cold dark matter
($\Lambda$CDM) 
cosmological parameters measured by the \cite{Planck2020}, including all effects available 
such as lensing, redshift space distortions (RSD), and nonlinear clustering to compute with the code 
{\tt CAMBsources}\footnote{\url{https://camb.info/sources/}} \citep{Challinor11} the 
fiducial angular power spectrum $C_\ell$ that was used as input.

This set of simulations was designed to be used in the angular analyses where we assumed 
a top-hat redshift bin ($0.105 \le z \le 0.115$) with a surface number density of $2.3$ galaxies 
{\it per} deg$^{2}$ (the same as in SDSS data). 
These simulated data share the observational features of the data set in the analyses, that is, 
the same number density and footprint sky area, and they were also weighted by the FKP scheme. 
The number of simulated cosmic objects in each catalog is $N_{sim} \simeq 16,000$. 
For this set of simulated maps, we adopted an angular resolution of $0.11\,\mbox{deg}^2$, given 
by the {\tt HEALPix}\footnote{\url{https://healpix.sourceforge.io/}}~\citep{Gorski05} parameter 
$N_{side} = 512$.

\section{Two-point correlation estimator}\label{sec3}

We performed a 2D BAO measurement at low redshift with the 
SDSS blue galaxy sample. 
This measurement complements similar 2D analyses performed following the same 
method applied to other cosmological tracers, such as LRG and 
quasars, at several redshifts~\citep{Carvalho16,Alcaniz,Edilson18,Carvalho20}. 
The 2D BAO studies were performed by applying the 2PACF to the thin redshift bin 
$0.105 \le z \le 0.115$. 
Additionally, supported by a large set of log-normal simulations, we describe how the 
covariance matrix was used in the error analyses, including statistical and systematic 
contributions.

\subsection{Two-point angular correlation function}\label{2-pointACF}

\noindent
In the 2D analysis, the 2PACF~\citep{PeeblesYu,Landy-Szalay} estimates the angular 
correlation for data pairs projected on the celestial sphere 
\citep[for alternative clustering analyses, see, e.g.,][]{Avila18,Avila19,Bengaly17,%
Feldbrugge19,Novaes16,Novaes18,Marques20a,Marques20b,Pandey20,Sosa20}. 
Considering blue galaxies in a redshift shell, the 2PACF measures the angular diameter 
distance $D_{\!A}$ due to the transverse BAO signal of the sound horizon scale there, where 
this signature appears as a bump at certain angular scale. 
The expression for the 2PACF estimator, $\omega(\theta)$, is given by 
\begin{eqnarray}\label{2PACF}
\omega(\theta) \equiv \frac{DD(\theta)-2DR(\theta)+RR(\theta)}{RR(\theta)} \, ,
\end{eqnarray}
with $\theta$ the angular separation between any pair of blue galaxies $A, B$, given by 
\begin{eqnarray}
\,\,\theta\,=\,\arccos[\sin{\delta_{A}}\sin{\delta_{B}}+
\cos{\delta_{A}}\cos{\delta_{B}}\cos(\alpha_{A}-\alpha_{B})] \, , \nonumber
\end{eqnarray}
where $\alpha_{A}, \alpha_{B}$ and $\delta_{A}, \delta_{B}$ are the right ascension and 
declination coordinates of the blue galaxies $A$ and $B$, 
respectively~\citep{Landy-Szalay,Sanchez11}.

To find the angular scale $\theta_{\mbox{\footnotesize\sc fit}}$ of the BAO bump in the 2PACF, 
$\omega(\theta)$, we used the method proposed by \cite{Sanchez11}, 
which is based on the empirical parameterization of $\omega = \omega(\theta)$, 
\begin{eqnarray}\label{eqFIT}
\omega(\theta) = A + B \, \theta^{\,\gamma} + C \exp^{-(\theta-\theta_{\rm FIT})^2 / 
2\sigma_{\rm FIT}^{2}} \, ,
\end{eqnarray}
where $A$, $B$, $\gamma$, $C$, $\theta_{\mbox{\footnotesize\sc fit}}$, and 
$\sigma_{\mbox{\footnotesize\sc fit}}$ are free parameters. 
Therefore this equation provides the BAO bump best-fit, 
$\theta_{\mbox{\footnotesize\sc fit}}$, and the width of the bump is 
$\sigma_{\mbox{\footnotesize\sc fit}}$. 

The final measurement of the acoustic peak was obtained after 
accounting for the shift due to the projection 
effect~\citep{Sanchez11,Carnero12,Carvalho16}.
In the 2D analyses, all galaxies in the redshift bin in the study with thickness 
$\delta z$ are assumed to be projected onto the celestial sphere. 
Thus, the finite thickness of the shell, $\delta z \ne 0$, produces a shift of the BAO peak. 
This shift is estimated through numerical analysis by assuming a fiducial cosmology 
(as we show in Sect.~\ref{sec4.2}), 
but the results show that for thin shells, the shift is small and weakly dependent on the 
cosmological parameters \citep[for details, see, e.g.,][]{Sanchez11,Carvalho16}. 
The redshift shell should be as thin as possible to minimize the projection effect that 
affects the measurement by erasing the acoustic signature, but at the same time, it should be 
a numerically dense data set, enough to obtain a good BAO S/N. 
We calculate the shift to be applied to $\theta_{\mbox{\footnotesize\sc fit}}$ due the projection effect in Sect.~\ref{sec4.2}.

\subsection{Covariance matrix estimation}\label{sec:covMat}

To estimate the covariance matrix and the significance of our results, we used 
the galaxy mocks described above (see Sect.~\ref{mocks}). 
For each mock, we extracted the 2PACF information from a set of $N_b$ bins in which the 
interval of $\theta$ values was divided. 
The covariance matrix for $\omega(\theta)$ was estimated using the expression 
\begin{eqnarray}\label{covmateq}
\mbox{Cov}_{ij} = 
\frac{1}{N}\sum_{k=1}^{N} 
      \Big[\mbox{w}_k(\theta_i)-\overline{\mbox{w}}(\theta_i)\Big]
      \Big[\mbox{w}_k(\theta_j)-\overline{\mbox{w}}(\theta_j)\Big]\,,
\end{eqnarray} 
where the $i$ and $j$ indices represent each $\theta$ bin, $i,j=1,\dots,N_b$, 
and $\mbox{w}_k$ is the 2PACF for the $k$-th mock catalog, with $k=1,\dots,N$; 
$\overline{\mbox{w}}(\theta_i)$ is the mean value for this statistics over the $N = 1,000$ 
mocks in that bin. 
Finally, the error of $\mbox{w}(\theta_i)$ is the square root of the main diagonal, 
$\Delta \mbox{w}(\theta_i) = \sqrt{\mbox{Cov}_{ii}}$.

\vspace{0.3cm}
\section{Clustering analyses in 2D}\label{sec4}

\begin{figure}
\hspace*{-0.2cm}
\includegraphics[scale=0.59]{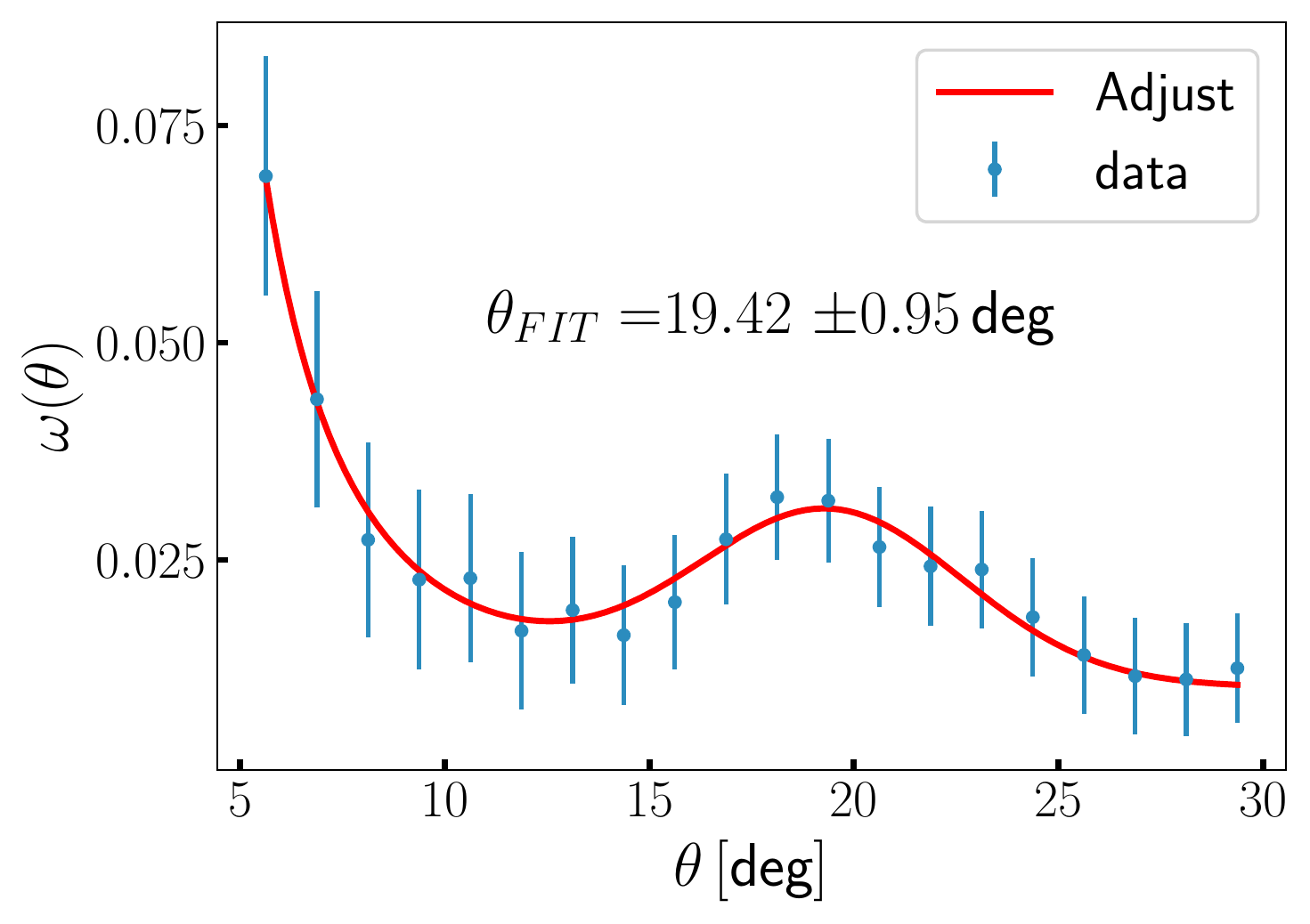}
\caption{BAO signature obtained in the 2PACF by analyzing the sample in the redshift interval 
$z \in [0.105, 0.115]$, with $\delta z = 0.01$. The bin size in this 2PACF is $1.25^{\circ}$, and we 
use 50 random catalogs with the same observational features as the galaxy catalog.}
\label{fig:2PACF}
\end{figure}

We studied the clustering of the SDSS blue galaxy sample performing 2D analysis. 
Using the estimator given by Eq. (\ref{2PACF}), $\omega(\theta)$, we 
calculated the 2PACF for our sample of $N_g = 15,942$ galaxies in the redshift interval 
$z \in [0.105,0.115]$, with effective redshift $z_{\mbox{\tiny eff}} = 0.11$. 
The 2PACF was calculated using \texttt{TREECORR}~\citep{Jarvis04} for equally spaced 
values of $\theta$ in the interval 
$5^{\circ} \leq \theta \leq {30^{\circ}}$, in a total of $\mbox{N}_{b}=20$ bins, which means 
that the bin size was $1.25^{\circ}$. 
To extract the BAO bump position, we used Eq.~(\ref{eqFIT}) to fit the 2PACF data 
through a least-squares method; the errors in the parameters correspond to the statistical uncertainties provided by the fitting procedure (the estimated covariance matrix of the parameters).
The result is shown in Fig. \ref{fig:2PACF}, where 
$\theta_{\mbox{\footnotesize\sc fit}} = 19.42^{\circ}$. 
Our result for this procedure is summarized in Table~\ref{table1}, where we display the best-fit 
parameters obtained in this fitting approach using Eq.~(\ref{eqFIT}). 

\linespread{1.25}
\begin{table}
\centering
\setlength{\extrarowheight}{0.12cm}
\begin{tabular}{|c|c|}
\hline
Parameters \,\,& Equation~(\ref{eqFIT}) 
\\ [0.05cm] \hline
$A$                    & \,\,$9.92\pm 6.41\,\,(\times10^{-3})$\\
$B$                    & \,\,$0.77\pm 2.05\,\,(\times10^{-4})$ \\ 
$\gamma$               & \,\,$2.86\pm 1.06$ \\
$C$                    & \,\,$19.29\pm 7.43\,\,(\times10^{-3})$ \\
$\sigma_{FIT}$         & \,\,~3.26$^\circ\pm$0.96$^\circ$ \\
$\theta_{FIT}$         & \,\,~~19.42$^\circ\pm$0.95$^\circ$ (stat) 
\\ [0.1cm] \hline
\end{tabular}
\vspace{0.2cm}
\caption{Best-fit parameters of Eq.~(\ref{eqFIT}), obtained through the $\chi^2$ statistics, 
Eq.~(\ref{covmatrix}), using the covariance matrix shown in Fig.~\ref{fig:matrizCov}.
}
\label{table1}
\end{table}

\subsection{Statistical significance}

\begin{figure}
\hspace*{0.1cm}
\includegraphics[scale=0.38]{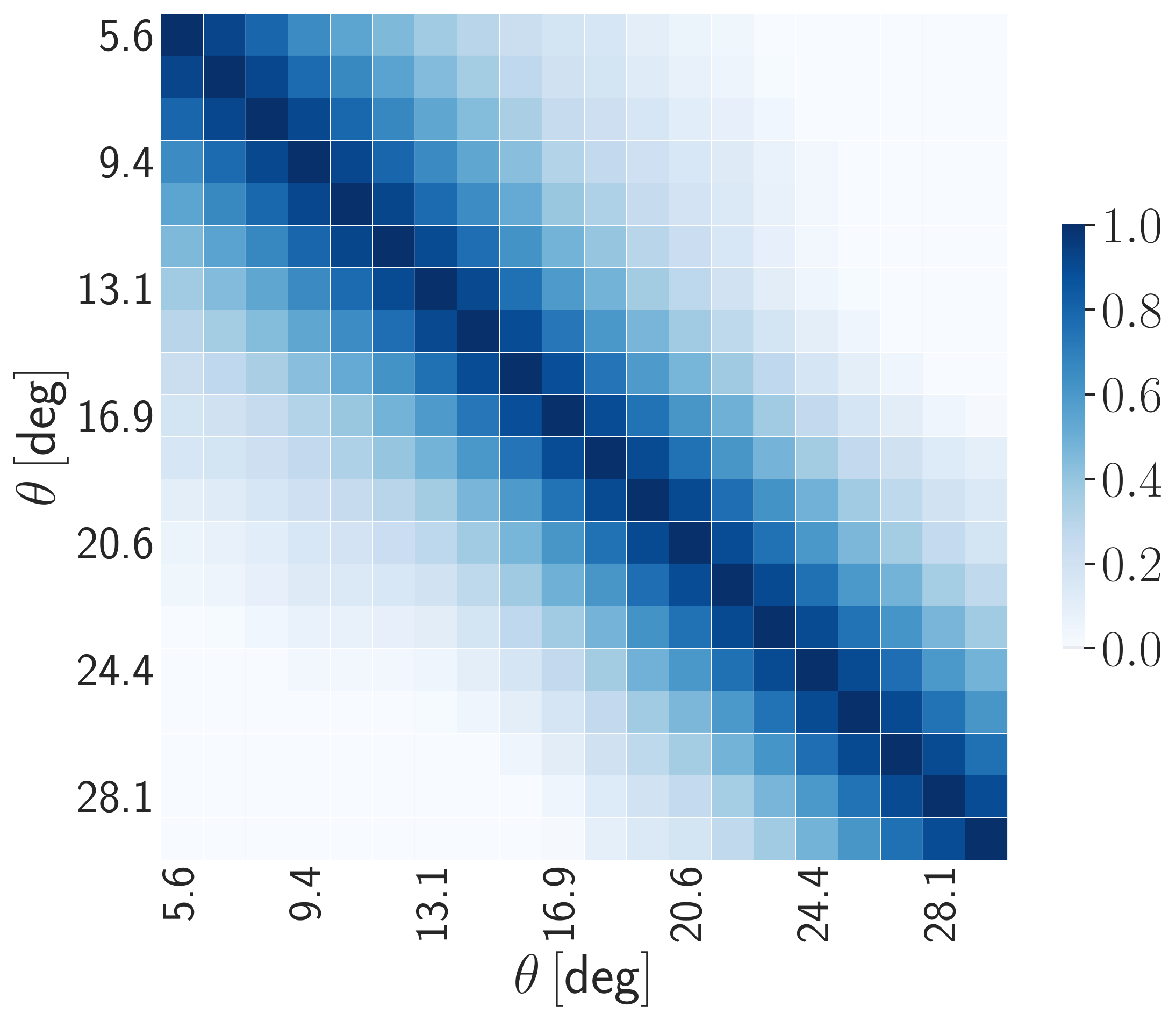}
\caption{Covariance matrix for the 2PACF obtained from Eq.~(\ref{covmateq}) using the 
set of log-normal simulated maps (see Sect.~\ref{mocks}).}
\label{fig:matrizCov}
\end{figure}

The statistical significance of the BAO angular measurement was obtained through the $\chi^2$ 
method, 
\begin{eqnarray}\label{covmatrix}
\chi^2(\alpha) = \left[\mbox{w} - \mbox{w}^{\mbox{\small\sc f\/it}}(\alpha)\right]^{T}\mbox{Cov}^{-1} 
\left[\mbox{w} - \mbox{w}^{\mbox{\small\sc f\/it}}(\alpha)\right] \, , 
\end{eqnarray} 
where we used the inverse of the covariance matrix, $\mbox{Cov}$, estimated as described 
in Sect. \ref{sec:covMat} and shown in Fig. \ref{fig:matrizCov}.
The symbols $\left[\,\right]$ and $\left[\,\right]^{T}$ represent column vectors and row 
vectors, respectively.

Following \cite{Edilson18}, we adjusted the parameters of 
Eq.~(\ref{eqFIT}) based on the minimum $\chi^2$ method for $\alpha \in [0.85, 1.25]$, 
which is called the scale dilation parameter, for two cases: 
considering $C$ as  free parameter, $C \ne 0$ ($\chi^{2}_{min}=13.13$), and imposing 
$C = 0$ ($\chi^{2}_{min}=22.06$), where $\chi^{2}_{min}$ corresponds to 
$\alpha_{min}=0.996$, the latter case representing the \textit{\textup{non}}-BAO case. 
Table~\ref{table1} shows the best-fit parameters obtained considering $\alpha=1$.

As a result, the best-fit\footnote{Here $\alpha$ is also accounted for as a free parameter, that is, we have a total of four and seven free parameters in the non-BAO and BAO cases, respectively.} 
of the {\it \textup{non}}-BAO case (16 degrees of freedom, {\it \textup{dof}}), compared to 
the BAO case (13 {\it \textup{dof}}), is disfavored by $\Delta \chi^2 = 8.93$. 
Therefore our BAO angular detection has a statistical significance of $2.2\,\sigma$, 
which is compatible with the distance of the $C$ parameter from zero.

\vspace{0.4cm}
\subsection{Projection effect in the 2PACF}\label{sec4.2}

\noindent
To know the angular scale $\theta_{\mbox{\sc bao}}$ , we need to correct the 
$\theta_{\mbox{\sc fit}}$ due to the projection effect, which produces a shift in the BAO bump 
position~\citep{Sanchez11}. 
To quantify this shift, we first computed the expected angular BAO scale, $\theta_{E}^{0}$, 
corresponding to the bump position for the case $\delta z=0$ calculated from the expected 
2PACF, 
\begin{eqnarray} \label{expected}
w_{E}(\theta, z) = \int_0^\infty dz_1 \,\phi(z_1) \int_0^\infty dz_2 \,\phi(z_2) 
\,\xi_{E}(s, z) \, , 
\end{eqnarray}
where $z = z_{\mbox{\tiny eff}} = (z_1 + z_2) / 2$, with $z_2 = z_1 + \delta z$, 
and $\phi(z_i)$ is the normalized galaxy selection function at redshift $z_i$.
The function $\xi_{E}$ is the 2PCF expected in the fiducial cosmology, given by 
\citep[see, e.g.,][]{Sanchez11}
\begin{eqnarray} \label{pktoxi}
\xi_{E}(s,z)=\! \int_0^\infty \! \frac{dk}{2\pi^2} \, k^2 \, j_0(k s) \, b^2 \, P_m(k, z) \, , 
\end{eqnarray}
where $j_0$ is the zeroth-order Bessel function, $P_m(k, z)$ is the matter power spectrum, and 
$b$ is the bias factor. 

It is suitable to examine the RSD effect on the 
measurement of the angular BAO signature. 
For this we performed analyses that included the linear RSD by changing $P_m(k,z)$ by 
$(1 + \beta \,\mu^2)^2 P_m(k,z)$ in the Eq.~(\ref{pktoxi}), where we considered 
two cases: 
the linear, $P_{m}^{L}(k,z)$, and the nonlinear, $P_{m}^{NL}(k,z)$, matter power spectra 
produced using the numerical code CAMB~\citep{Challinor11}, at $z=0.11$. 
We assumed the $\Lambda$CDM model with the cosmological parameters 
measured by the~\cite{Planck2020}. 
The term $(1 + \beta \,\mu^2)^2$ corresponds to the Kaiser model for large-scale 
RSD~\citep{Kaiser87}, where $\beta$, the velocity scale parameter is $\beta = f / b$, $b$ 
is the linear bias, and $f$ is the growth rate of cosmic structures, with 
$f \simeq \Omega_m(z)^{0.55}$, where $\mu$ is the cosine of the angle between the wave 
vector $\mbox{\bf k}$ and the line of sight. 

Our results show that the relative difference in the bump position between the linear matter 
power spectra with and without the linear RSD is $0.84\,\%$. 
For the nonlinear matter power spectra with and without the linear RSD, the relative 
difference in the bump position is equal to $0.70\,\%$. 
On the other hand, comparing the linear and the nonlinear matter power spectra cases, 
the differences in the bump position are $0.14\,\%$ and $0.00\,\%$ for the cases with and 
without the linear RSD effect, respectively. 
In all cases the relative differences are smaller than $1\,\%$, therefore we conclude that these effects on our angular BAO measurement are small and are included in the 
final error.

Next, we applied this procedure to our data, where $\delta z = 0.01$ is the thickness of the redshift bin used here, to find $\theta_{E}^{\delta z}$. 
Then, the BAO angular scale, $\theta_{\mbox{\sc bao}}$, is  
\begin{eqnarray}
\theta_{\mbox{\sc bao}}(z) \,=\, \theta_{\mbox{\footnotesize\sc fit}}(z) 
+ \Delta \theta(z, \delta z) \, \theta_{\mbox{\footnotesize\sc fit}}(z) \, , 
\end{eqnarray}
where $\Delta \theta(z, \delta z) \equiv (\theta_{E}^{0} - \theta_{E}^{\delta z})/\theta_{E}^{0}$ 
shifts the fitted value $\theta_{\mbox{\footnotesize\sc fit}}(z)$ to the correct acoustic 
scale $\theta_{\mbox{\sc bao}}(z)$, at $z=z_{\mbox{\tiny eff}} = 0.11$.

Assuming the $\Lambda$CDM model with the cosmological parameters measured 
by~\cite{Planck2020},
we estimate $\theta^0_E=17.93^\circ$ and $\theta^{0.01}_E=17.58^\circ$, which 
corresponds to $\Delta\theta=1.96\%$. 
Consequently, $\theta_{\mbox{\sc bao}}(z_{\mbox{\tiny eff}} = 0.11) = 19.8^\circ$. 
As shown by \cite{Sanchez11}, the choice of the fiducial cosmological model introduces a 
systematic error of $1\%$ in the final $\theta_{\mbox{\sc bao}}$ error.

\subsection{Robustness of the BAO signal}\label{robust}

\noindent
We performed a robustness test in the two-point angular correlation statistics to confirm the BAO 
signature in the 2PACF. To verify that the BAO signature corresponds to a robust detection, we performed the 
{\it \textup{small-shifts criterion}} test~\citep{Carvalho16, Edilson18}. 
The main idea here is to distinguish between the true BAO bump, which is expected to be 
smoothed, but survives, under weak perturbations in the galaxy positions, while other local 
maxima that originate in systematic effects or statistical noise tend to disappear in a reanalysis 
after the perturbations. 
For this, we first generated 100 modified galaxy catalogs by drawing the modified 
position of each galaxy resulting from a random Gaussian distribution with the mean equal to the original position 
and standard deviation $\sigma_s$, and for each modified catalog we calculated the 2PACF curve. 
The final 2PACF was estimated as the average over the 100 curves resulting from each of the 
modified galaxy catalogs, perturbed considering the standard deviation $\sigma_s$. 
We performed this process for the cases with $\sigma_s=1.0^{\circ}, 2.0^{\circ}$, and $3.0^{\circ}$. 
In the calculation of each 2PACF we used the same set of 50 random catalogs as in the 
main analysis, always applying Eq.~(\ref{2PACF}).

\begin{figure}
\hspace*{-0.4cm}
\includegraphics[scale=0.6]{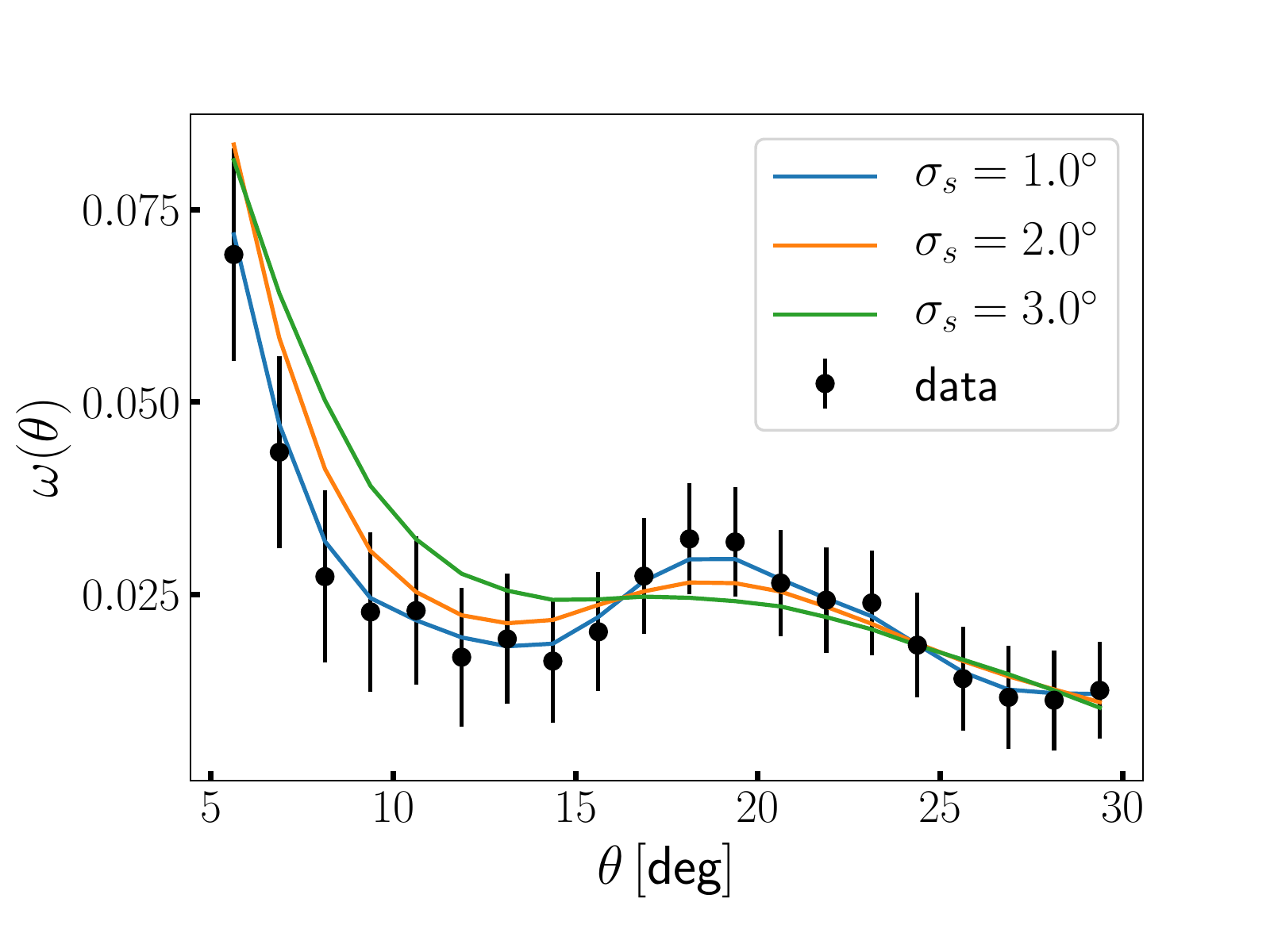}
\caption{Robustness test analyses. 
We performed small random shifts in the galaxies angular coordinates and repeated the 2PACF calculation.
The black dots represent the original data analysis, and the curves correspond to the cases 
we studied (as indicated in the legend).}
\label{fig:randomShifts}
\end{figure}

Our results are shown in Fig. \ref{fig:randomShifts}, where we present the original data as 
black dots together with the three cases for $\sigma_s$ that are printed as solid colored curves. 
As observed, the larger the random displacements in the blue galaxy angular positions, the 
smoother the 2PACF curves. This also smoothes the BAO bump signature. 
Simultaneously, these displacements also smooth other maxima and minima that may originate 
from systematic effects or statistical noise and appear in the original 2PACF.

\subsection{Spectroscopic-z error}

As shown by \cite{Sanchez11}, the main source of error in the BAO signal for photometric 
surveys is the uncertainty in the measurement of the redshift, $z$. 
Although we study spectroscopic data, for which this uncertainty is smaller, 
it is important to quantify this source of error in the final BAO measurement. 
To do this, we constructed 300 spec-$z$ simulations for which we 
considered the measured $z$ of each blue galaxy as the true one plus a random error obtained 
from a Gaussian distribution of zero mean and standard deviation given by its measured 
uncertainty that is available in the data catalog \citep[see][]{Edilson20}.

\begin{figure}
\hspace*{-0.4cm}
\includegraphics[scale=0.6]{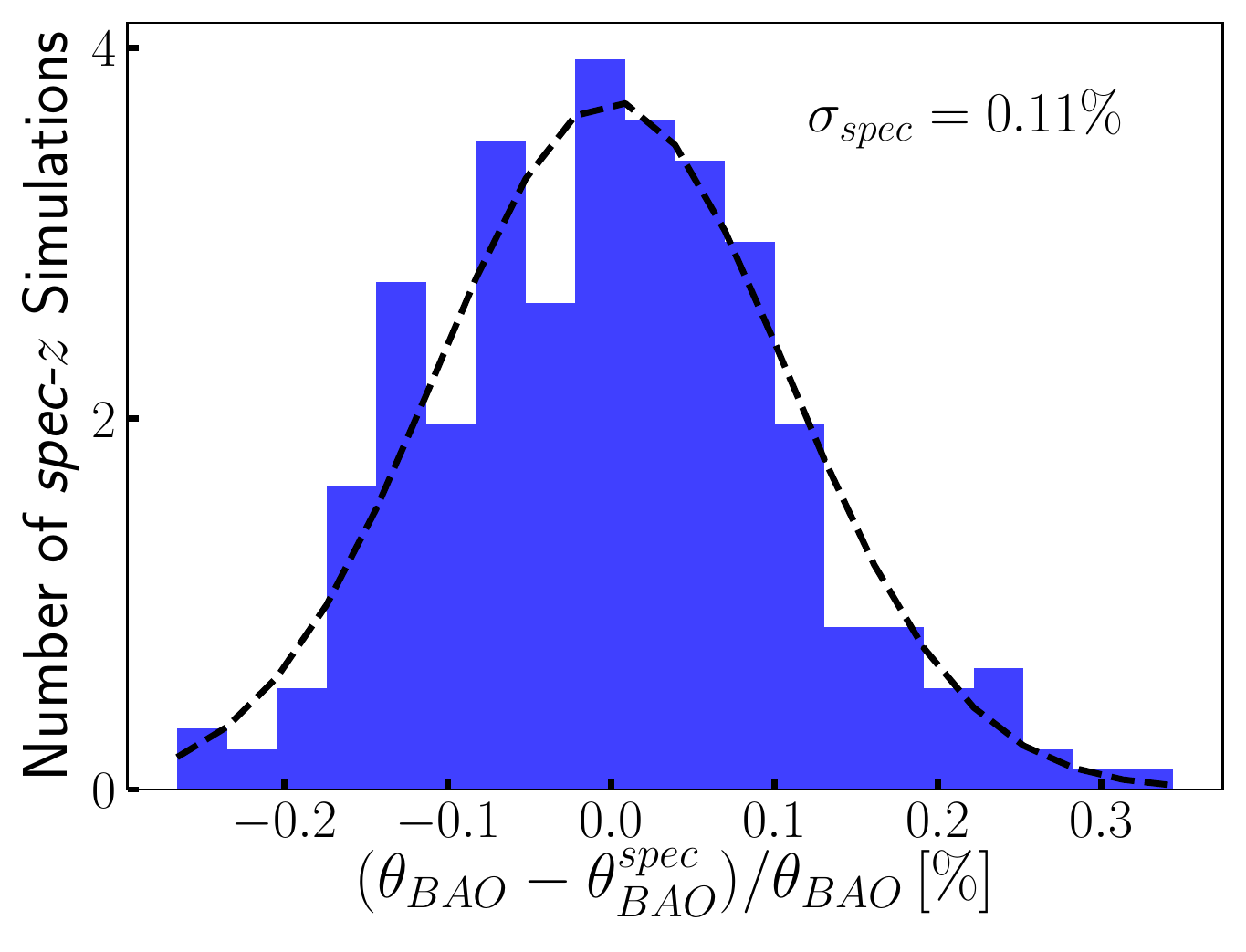}
\caption{Histogram of the relative difference, in percentage ($\%$), between 
the BAO scale obtained from the blue galaxy data, $\theta_{\mbox{\sc bao}}$, and those 
obtained from each simulated spec-$z$ catalog, $\theta_{BAO}^{spec}$. As expected, for 
the spectroscopic data the z errors affect the measurements of the BAO 
signature little. They contribute to the final error with only $0.11\%$.}
\label{fig:HistBAO}
\end{figure}

The results of this analysis are displayed in Fig. \ref{fig:HistBAO}. 
There we show the histogram of the relative difference between the $\theta_{\mbox{\sc fit}}$ 
measured for each spec-$z$ simulation and the $\theta_{\mbox{\sc fit}}$ from the blue 
galaxy data. 
As expected, in the case of spectroscopic data as for the blue galaxy data analyzed here, 
the error coming from the $z$-uncertainty introduces a small error in the final angular 
BAO measurement of $0.11\%$.

The 2D BAO measurement performed here complements a set of other measurements 
obtained with the same method~\citep{Carvalho16,Alcaniz,Edilson18,Carvalho20}. 
Thus, our angular BAO measurement at $z_{\mbox{\tiny eff}} = 0.11$ is 
$\theta_{\mbox{\sc bao}} = 19.8^\circ \pm 1.05^\circ$. 
This error in the $\theta_{\mbox{\sc bao}}$ measurement includes 
the statistical and systematic errors due to the spectroscopic-z error, parameterization,
RSD, projection effect, and nonlinearities \citep[see][for a broader 
discussion of the error estimation]{Sanchez11}.

\subsection{Cosmological constraints from $\theta_{\mbox{\sc bao}}$ data}\label{cosmocons}

\begin{figure}
\hspace*{-0.3cm}
\includegraphics[scale=0.6]{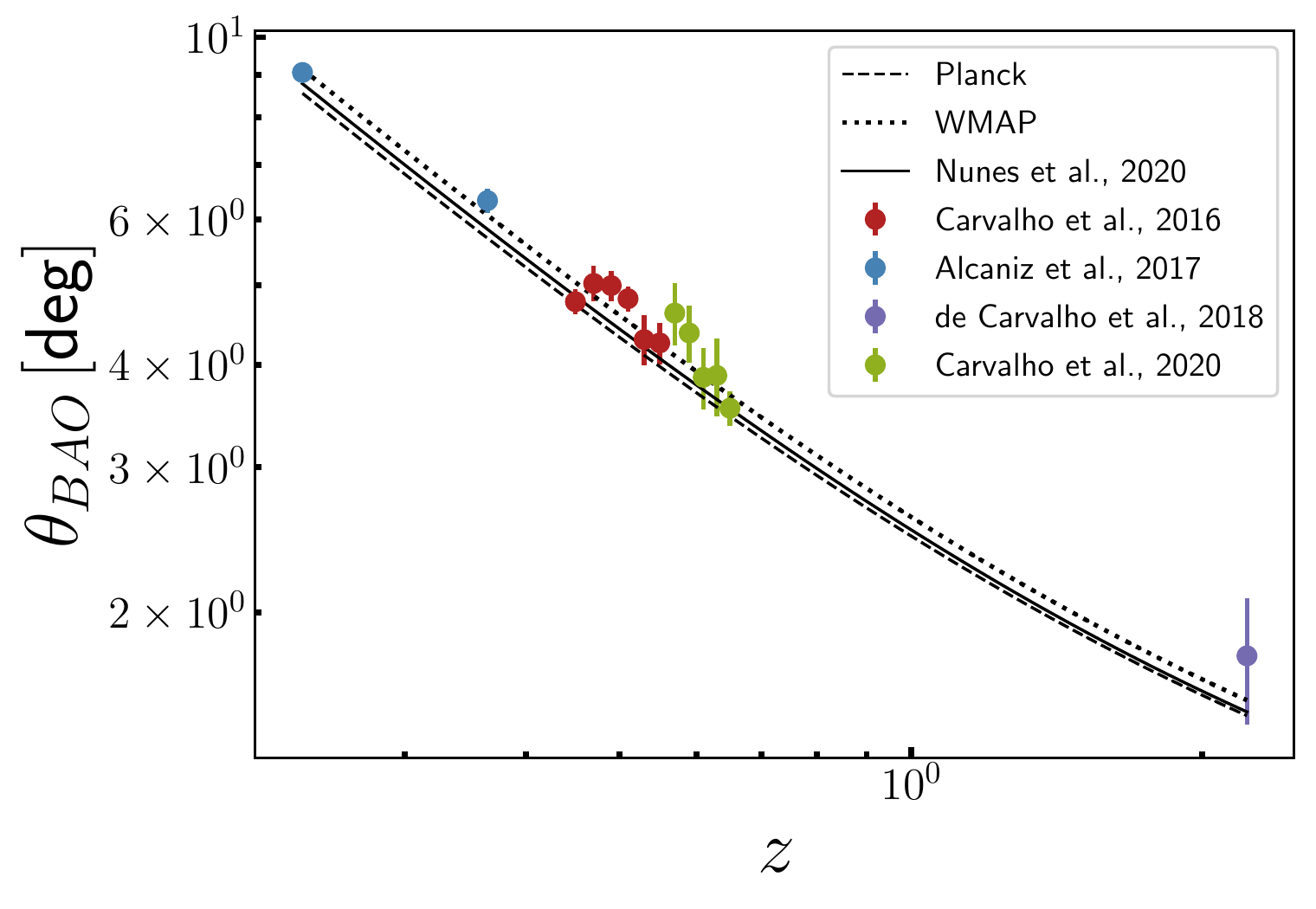}
\caption{$\theta_{\mbox{\sc bao}}$ measurements as a function of redshift. 
When the $\Lambda$CDM model is assumed, the dashed, dotted, and continuous lines represent 
the results obtained using the cosmological parameters from~\cite{Planck2020}, 
\cite{wmap9} (WMAP team), and~\cite{Nunes1}, respectively.
}
\label{fig:thetas}
\end{figure}

Here we combined our 2D measurement of the BAO scale at low redshift, 
$z_{\mbox{\tiny eff}} = 0.11$, with the results obtained 
by \cite{Carvalho16}, \cite{Alcaniz}, \cite{Edilson18} and \cite{Carvalho20} (see Fig.~\ref{fig:thetas}), to constrain the parameters of the 
$\Lambda$CDM, $w$CDM, and $w(t)$CDM models. 
For the $w(t)$CDM model we chose the Barboza-Alcaniz 
parameterization~\citep{Barboza08}. 
To be consistent with the error determination from the analyses mentioned 
above and to allow a proper combination of the results, in this section we adopt 
$\sigma_{\mbox{\sc bao}}=\sigma_{\mbox{\footnotesize\sc fit}}=3.26^\circ$ 
as the error in the measurement of $\theta_{\mbox{\sc bao}}$, that is, 
$\theta_{\mbox{\sc bao}}(0.11)=19.8^\circ \pm 3.26^\circ$.

\begin{figure}
\hspace*{-0.4cm}\includegraphics[scale=0.49]{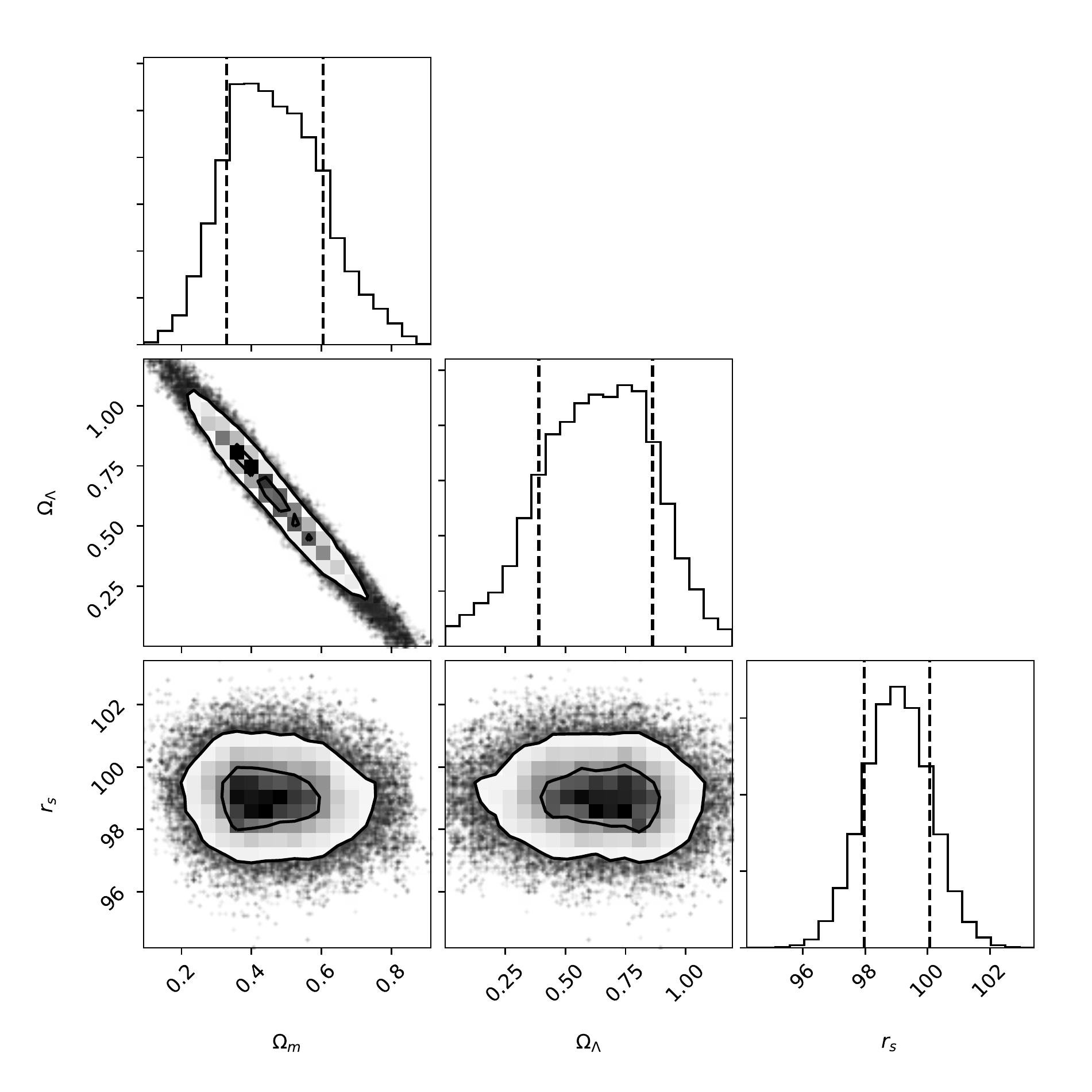}
\caption{Contour levels at $1\,\sigma$ (inner) and $2\,\sigma$ (outer) for
the $\Omega_m$, $\Omega_\Lambda$ and $r_s$ parameters extracted from the BAO angular 
scale data and the $r_s$ prior defined from the Planck value of~\cite{Planck2020}, namely 
$r_s = 99.08 \pm 0.18\,h^{-1}$Mpc. 
In the histogram plots, the dashed vertical lines represent the $\pm 1\,\sigma$ values.}
\label{fig:triangle1_BG_100000}
\end{figure}

To restrict the cosmological parameters of the models $\Lambda$CDM, $w$CDM, and 
$w(t)$CDM, we performed Markov chain Monte Carlo (MCMC) analyses to explore the 
parameter space. 
In all cases we assumed $\Omega_k=0$. 
The analyses were performed using the code 
{\tt PyMC}\footnote{\url{https://pymc-devs.github.io/pymc}}, 
assuming uniform priors for all the parameters in study except for $r_s$, for which we assumed a 
Gaussian prior with a standard deviation equal to the measurement error. 
We investigated three different values of $r_s$: 
$r_s = 99.08 \pm 0.18\,h^{-1}$Mpc obtained by the \cite{Planck2020}, 
$r_s = 106.61 \pm 3.47\,h^{-1}$Mpc calculated by \cite{wmap9} (WMAP team), 
and $r_s = 102.2 \pm 0.2\,h^{-1}$Mpc calculated by~\cite{Nunes1}. 
The results of these analyses are displayed in Table~\ref{table2}, where the uncertainties 
correspond to $1\,\sigma$ errors. 
Figure~\ref{fig:triangle1_BG_100000}\, 
shows the constraints for 
$\Lambda$CDM model case as an example. The inner and the outer curves represent the 
$1\,\sigma$ and $2\,\sigma$ contour levels, respectively
\footnote{This figure was made 
using the~\texttt{Corner Plot} software developed by \cite{Foreman}.}. 
In the histogram plots, the $\pm 1\,\sigma$ values are shown as dashed vertical lines. 
In all cases, the data prefer a high fraction of matter, with $\Omega_m > 0.4$.
Moreover, the results from the $w$CDM and $w(t)$CDM models
are consistent with the $\Lambda$CDM model, although the 
analyses reveal a poorer constraint of the parameters. 

In general, we observe that the results are consistent with each other, but when the value from the Planck Collaboration, 
$r_s = 99.08 \pm 0.18\,h^{-1}$Mpc, is used 
as a prior for $r_s$  , the constraints show the largest error bars (see 
Table~\ref{table2}). According to our tests, they are probably due to the small error of 
this parameter.

\linespread{1.25}
\begin{table*}
\centering
\scalebox{1.1}{
\setlength{\extrarowheight}{0.1cm}
\begin{tabular}{|c|c|c|c|c|}
\hline
\,\,model\,\,          & \,\,\,parameters\,\,\, & \,\,\,  $r_s^{Planck}$ 
\,\,\, & \,\,\, $r_s^{WMAP}$ \,\,\, &  \,\,\,$r_s^{Nunes\,et\,al.\,(2020a)}$
\\ [0.1cm] \hline 

{$\Lambda$CDM}   & $\Omega_m$       & 
$0.481\pm 0.140$   & $0.408\pm 0.115$ & $0.448\pm 0.128$ \\
                                & $\Omega_\Lambda$ & 
$0.608\pm 0.233$   & $0.532\pm 0.194$ & $0.578\pm 0.217$\\
                                & $r_s$            & 
$98.858\pm 2.362$ & \,\,\,$106.605\pm 0.537$\,\, & \,\,\,$102.014\pm 2.227$ \\
                                & $\chi^2_{\mbox{\tiny{reduced}}}$ & 
$1.026$ & $1.028$  &  $1.027$
\\  [0.1cm] \hline

{$w$CDM}           & $\Omega_m$       & 
$0.469\pm 0.157$ & $0.417\pm 0.125$ &  $0.427\pm 0.153$ \\
                             & $w$  & 
$-1.083\pm 0.499$ & $-1.152\pm 0.484$ &  $-1.079\pm 0.488$ \\
                             & $r_s$            & 
$98.961\pm 2.301$ & \,\,\,$106.597\pm 0.540$ & \,\,\, $102.002\pm 2.176$ \\   
                             & $\chi^2_{\mbox{\tiny{reduced}}}$ & 
$1.186$ & $1.170$ &  $1.191$
\\  [0.1cm] \hline

{$w(t)$CDM} & $w_0$            & 
$-0.855\pm 0.368$ & $-1.194\pm 0.244$ & $-1.000\pm 0.321$ \\
                                 & $w_1$            & 
$-0.111\pm 0.562$ & $-0.012\pm 0.573$ &  $-0.059\pm 0.569$ \\ 
                                 & $r_s$            & 
$99.253\pm 2.187$ & \,\,\,$106.556\pm 0.533$ & \,\,\, $101.980\pm 2.100$ \\  
                                 & $\chi^2_{\mbox{\tiny{reduced}}}$ & 
$1.247$ & $1.136$ &  $1.167$
\\ [0.1cm] \hline
\end{tabular}
}
\vspace{0.4cm}
\caption{
Constrained parameters for the models $\Lambda$CDM, $w$CDM, and $w(t)$CDM 
considering the sound horizon scale, $r_s$, given by the \cite{Planck2020}, 
WMAP \cite{wmap9}, and~\cite{Nunes1}. 
The units for the values of $r_s$ are Mpc$/h$.
}
\label{table2}
\end{table*}

Finally, we calculated the angular diameter distance at $z_{\mbox{\tiny eff}} = 0.11$.
The angular scale of the BAO bump, $\theta_{\mbox{\sc bao}}(z)$, is related to the angular 
diameter distance $D_A(z)$ and the sound horizon $r_s$ by 
\begin{eqnarray} 
D_A(z) \,=\, \frac{r_s}{(1 + z) \, \theta_{\mbox{\sc bao}}(z)} \,\, .
\label{thetaz}
\end{eqnarray}
When we use our angular BAO measurement, 
$\theta_{\mbox{\sc bao}}(0.11) = 19.8^\circ \pm 1.05^\circ$, 
and the sound horizon scale calculated by the~\cite{Planck2020}, 
$r_s = 99.08 \pm 0.18\,h^{-1}$Mpc, we use Eq.~(\ref{thetaz}) to determine the angular 
diameter distance at $z_{\mbox{\tiny eff}} = 0.11$:\, 
$D_A(0.11) = 258.31 \pm 13.71\, h^{-1}$Mpc.

\section{Conclusions}\label{sec5}

The availability of a rich set of astronomical data, 
mapping diverse cosmological tracers in various redshift intervals observed in large sky 
regions during long-time surveys makes these times exciting. 
This motivated us to study the BAO phenomenon in the local Universe, performing 2D 
statistical analyses of the low-redshift blue galaxy sample from the SDSS-DR12.

In the 2D clustering analyses performed here, we used a sample containing \,$N_g = 15,942$\, 
blue galaxies in the thin redshift bin $z \in [0.105, 0.115]$ with $z_{\mbox{\tiny eff}} = 0.11$. 
Applying the 2PACF estimator on these data, we measured the transverse 
BAO signature at \,$\theta_{\mbox{\sc bao}}(0.11)\,=\, 19.8^\circ\pm 1.05^\circ$ with a statistical significance of 
\,$2.2\,\sigma$\, . 
Our error analyses included statistical and systematic contributions. 
We also performed analyses that confirm the robustness of this transverse BAO measurement 
(see Sect.~\ref{robust} for details).

Additionally, we used the sound horizon scale calculated by the~\cite{Planck2020}, 
$r_s = 99.08 \pm 0.18\,h^{-1}$Mpc, to obtain a measurement of the angular diameter 
distance. 
Using this value of $r_s$ and our result for the BAO angular scale 
$\theta_{\mbox{\sc bao}}(0.11)$ in Eq.~(\ref{thetaz}), 
we obtained a measurement of the angular diameter distance at 
$z_{\mbox{\tiny eff}} = 0.11$: $D_A(0.11) = 258.31 \pm 13.71\, h^{-1}$Mpc. 
For the error analyses, which requires the computation of the covariance matrices, 
we used a set of log-normal simulations with similar observational features as the SDSS data 
we analyzed, including effects such as lensing, RSD, and nonlinear clustering 
(see Sect.~\ref{mocks} for details).

Our measurement of $D_A(0.11)$ was obtained by applying a method that is weakly 
dependent on a cosmological model, exactly as other measurements of $D_A(z)$ were 
obtained~\citep{Carvalho16,Alcaniz,Edilson18,Carvalho20}.
Here we used these measurements to perform MCMC analyses that constrain cosmological 
parameters of some $\Lambda$CDM-type models, namely $\Lambda$CDM itself, 
$w$CDM, and $w(t)$CDM. The results we obtained are shown in Table~\ref{table2} and 
Fig.~\ref{fig:triangle1_BG_100000}. They agree with the reported values for these 
models~\citep[for analyses of cosmological parameters, see, e.g.,][]{wmap9,Planck2020,%
Marques19,Nunes1,Nunes2}.

\begin{acknowledgements}
The authors thank the Brazilian agencies PROPG-CAPES/FAPEAM program, CNPq, 
CAPES, and FAPESP for the grants under which this work was carried out. 
\end{acknowledgements}


\end{document}